\documentclass[twocolumn,aps,pra,superscriptaddress,showpacs
,eqsecnum
]{revtex4}
\usepackage{bm,amsmath,amsfonts,amssymb,esint,graphicx,color}

\newcommand{\Eq}[1]{Eq.~\eqref{#1}}
\newcommand{\eq}[1]{\eqref{#1}}

\newcommand{\beq}{\begin{equation}}
\newcommand{\eeq}{\end{equation}}
\newcommand{\beqa}{\begin{eqnarray}}
\newcommand{\eeqa}{\end{eqnarray}}
\newcommand{\Beqa}{\begin{eqnarray*}}
\newcommand{\Eeqa}{\end{eqnarray*}}

\newcommand{\past}{{\phantom{\ast}}}
\newcommand{\msk}{\mkern 2mu}
\newcommand{\nmsk}{\mkern -2mu}

\def\XXint#1#2#3{{\setbox0=\hbox{$#1{#2#3}{\int}$}
     \vcenter{\hbox{$#2#3$}}\kern-.5\wd0}}

\begin{document}

\title{Effective mass of elementary excitations in Galilean-invariant integrable models}

\author{K. A. Matveev}
\affiliation{Materials Science Division, Argonne National Laboratory, Argonne, Illinois 60439, USA}
\author{M. Pustilnik}
\affiliation{School of Physics, Georgia Institute of Technology, Atlanta, Georgia 30332, USA
}

\begin{abstract}
We study low-energy excitations of one-dimensional Galilean-invariant models integrable by Bethe ansatz and characterized by nonsingular two-particle scattering phase shifts. We prove that the curvature of the excitation spectra is described by the recently proposed phenomenological expression for the effective mass. Our results apply to such models as the repulsive Lieb-Liniger model and the hyperbolic Calogero-Sutherland model.
\end{abstract}


\maketitle

\section{Introduction}
\label{intro}

The concept of elementary excitations~\cite{PinesNozieres} plays a fundamental role in our understanding of low-temperature physics of quantum systems. Although the interaction between the constituent particles in these systems may be strong, long-lived weakly interacting elementary excitations often suffice to completely describe the low-lying excited states. 
In one dimension, the nature and properties of such elementary excitations can be studied in the framework of a semi-phenomenological hydrodynamic description~\cite{hydrodynamics,Haldane_LL,Giamarchi}, similar to that originally developed in the theory of superfluidity~\cite{PinesNozieres}. 

In this paper we consider single component Galilean-invariant one-dimensional quantum liquids. 
For such systems, the hydrodynamic theory~\cite{hydrodynamics,Haldane_LL,Giamarchi} predicts the existence of two branches of elementary excitations with spectra~\cite{Rozhkov} 
\beq
\varepsilon_\pm(p) = v|p| \pm \frac{p^2\msk}{2 m_\ast} +\ldots,
\label{1.1}
\eeq
parametrized by the velocity $v$ and the effective mass $m_\ast$. The velocity is given by~\cite{PinesNozieres,Giamarchi}
\beq
v = \sqrt{\frac{n_0}{m\msk} \frac{d\mu\,}{\,dn_0}\msk},
\label{1.2}
\eeq 
where $n_0$ is the mean density of the constituent particles, $m$ is their mass, and $\mu$ is the chemical potential. The effective mass satisfies~\cite{Pereira,ISG} 
\beq
\frac{\,m_\past}{\,m_\ast} = \frac{1}{2v\sqrt{K}}\frac{d (vn_0)}{dn_0}.
\label{1.3}
\eeq
The Luttinger-liquid parameter $K$ in \Eq{1.3} is defined~\cite{Giamarchi} as \mbox{$K = v_F/v$}, where $v_F = \pi\hbar{\mkern 1mu} n_0/m$. (The notation reflects the fact that for noninteracting fermions $v_F$ coincides with the Fermi velocity.)

Although in principle the phenomenological expressions~\eq{1.1}-\eq{1.3} are expected to hold for a wide range of one-dimensional systems, their applicability to any given system is by no means guaranteed. In fact, \Eq{1.1} fails for the Calogero-Sutherland model in which the interaction potential falls off as inverse square of the distance between particles~\cite{Sutherland_book}. In this exactly solvable model, the spectra of the two excitation branches are characterized by different effective masses. This feature originates in the singular behavior~\cite{Sutherland_book} of the two-particle scattering phase shift. Thus, it is likely that the Calogero-Sutherland model represents an exception rather than the rule. It is therefore important to check the applicability of Eqs.~\eq{1.1}-\eq{1.3} to other exactly solvable models.

A relation for the effective mass similar to \Eq{1.3} holds for an antiferromagnetic spin chain in a magnetic field~\cite{Pereira}. In the case of Galilean-invariant systems, the validity of Eqs.~\eq{1.1}-\eq{1.3} is supported by the available in the literature results for the repulsive Lieb-Liniger model~\cite{Lieb,KBI,Sutherland_book} and for the hyperbolic Calogero-Sutherland model~\cite{Sutherland_book,Calogero,Sutherland}. The validity of the phenomenological expression \eq{1.2} for the velocity was proven for the Lieb-Liniger model in Ref.~\cite{Lieb}; the proof in fact applies~\cite{Sutherland_book} to both models. For the Lieb-Liniger model, the result \eq{1.3} for the effective mass was verified analytically for both very weak~\cite{PM_LiebLiniger,PM_KdV} and very strong~\cite{Zoran} repulsion, and confirmed numerically~\cite{AP} for any repulsion strength. For the hyperbolic Calogero-Sutherland model, \Eq{1.3} was shown to hold in the limit of a strong short-range repulsion~\cite{PM_WignerToda,PM_KdV}, when the model is equivalent~\cite{Sutherland,Sutherland_book,PM_KdV} to the quantum Toda model~\cite{qToda}.

In this paper, we demonstrate the applicability of Eqs.~\eq{1.1}-\eq{1.3} to Galilean-invariant Bethe ansatz-integrable models with nonsingular scattering phase shifts, including both the repulsive Lieb-Liniger model and the hyperbolic Calogero-Sutherland model, regardless of the choice of their parameters. In Sec.~\ref{phenomenology} we review the phenomenology of one-dimensional quantum liquids, including the derivation of Eqs.~\eq{1.1}-\eq{1.3}. In Sec.~\ref{Bethe} we use Bethe ansatz to evaluate the low-energy excitation spectra and prove the validity of \Eq{1.3}. The results are discussed in Sec.~\ref{Discussion}.

\vspace{-1.5ex}
\section{One-dimensional quantum liquid}
\label{phenomenology}

The hydrodynamic description~\cite{hydrodynamics,Haldane_LL,Giamarchi} of a one-dimensional quantum liquid is formulated in terms of two bosonic fields, $\varphi(x)$ and $\vartheta(x)$, obeying the commutation relations \mbox{$[\partial_x\varphi,\vartheta(y)] = -i\pi\delta(x-y)$} and \mbox{$[\varphi(x),\varphi(y)] = [\vartheta(x),\vartheta(y)] = 0$}. In terms of these fields, the particle density $n(x)$ and momentum per particle~$\kappa(x)$ are given by~\cite{Giamarchi,MA} 
\beq
n(x) = n_0 + \frac{1}{\pi}{\mkern 1mu}\partial_x\varphi,
\quad
\kappa(x) = - \msk\hbar\msk\partial_x\vartheta,
\label{2.1}
\eeq 
The field $\varphi$ obeys the periodic boundary condition \mbox{$\varphi(x + L) = \varphi(x)$}, where $L$ is the size of the system. We impose a similar condition on the field $\vartheta$ as well, \mbox{$\vartheta(x + L) = \vartheta(x)$}, which restricts our consideration to excitations near the zero-momentum ground state.

The total momentum of the liquid reads
\beq
P = \int_0^L\!dx\,n(x){\mkern 1mu}\kappa(x)
= -\,\frac{\hbar}{\pi}\!\int_0^L\!dx\,(\partial_x\varphi)(\partial_x\vartheta),
\label{2.2}
\eeq 
and the Hamiltonian is given by
\beq
H = \frac{1}{2m\msk}\!\int_0^L\!dx\msk n(x){\mkern 1mu}\kappa^2(x) 
+ U[{\mkern 1mu} n {\mkern 1mu}].
\label{2.3}
\eeq
The first term in the right-hand side of \Eq{2.3} represents the kinetic energy of the liquid. It has the universal form set by the Galilean invariance. On the other hand, $ U[{\mkern 1mu} n{\mkern 1mu}]$ is a model-dependent functional of density that depends on interactions. 

For low-energy excitations, deviations of the density from its mean are small, and the Hamiltonian can be expanded in $\partial_x\varphi$ and higher-order derivatives of~$\varphi$. With $\varphi\msk$- and $\vartheta\msk$-independent constant dropped, the leading contribution in the resulting gradient expansion includes operators of scaling dimension 2, 
\beq
H_0 = \frac{\msk\hbar v_F}{2\pi\msk}
\!\int_0^L\!dx\bigl[(\partial_x\vartheta)^2 + \alpha(\partial_x\varphi)^2\bigr].
\label{2.4}
\eeq
The first term in \Eq{2.4} comes from the kinetic energy, and its coefficient is universal. The model-dependent coefficient $\alpha$ in the second term can be related to the velocity of the excitations~\cite{Giamarchi}. Indeed, Heisenberg equations of motion \mbox{$\partial_t\varphi = -\msk v_F{\mkern 1mu}\partial_x\vartheta$} and \mbox{$\partial_t\vartheta = -\msk\alpha v_F{\mkern 1mu}\partial_x\varphi$} show that both fields propagate with velocity $v = \alpha^{1/2\msk}v_F$. This gives $\alpha = K^{-2}$ and brings \Eq{2.4} into the standard Luttinger-liquid form~\cite{Haldane_LL,Giamarchi,ISG}
\beq
H_0  = \frac{\hbar v}{2\pi}\!\int_0^L\!dx\bigl[K(\partial_x\vartheta)^2 + K^{-1}(\partial_x\varphi)^2\bigr].
\label{2.5}
\eeq
The velocity can be found by considering the change of the ground state energy per length $E_0$ caused by a small change of density \mbox{$n_0\to n_0 + \delta n_0$}, which amounts to the shift \mbox{$\partial_x\varphi \to \partial_x\varphi \msk+\msk \pi\delta n_0$}. Substitution into \Eq{2.5} shows that $\delta E_0 = \frac{1}{2} v^2 (m/n_0)(\delta n_0)^2$, leading to \Eq{1.2} with $\mu = dE_0/dn_0$.

The next contribution in the gradient expansion contains operators of scaling dimension 3,
\beq
H_1 = \frac{\,\hbar^2}{2\pi m\msk}\!\int_0^L\!dx\bigl[(\partial_x\varphi)(\partial_x\vartheta)^2 
+ \beta(\partial_x\varphi)^3\bigr].
\label{2.6}
\eeq 
Similar to \Eq{2.4}, the first term in \Eq{2.6} comes from the kinetic energy and is universal,  whereas the model-dependent coefficient $\beta$ in the second term can be expressed in terms of macroscopic parameters~\cite{ISG,MA} by considering a change of density \mbox{$n_0\to n_0 + \delta n_0$}. The change leads to a first-order in $\delta n_0$ correction to the coefficient of the second term in \Eq{2.5}, yielding the relation 
\beq
\beta = \frac{\msk m}{3\pi\hbar}\frac{d(v/K)}{\msk dn_0}.
\label{2.7}
\eeq

At this point, it is convenient to switch from $\varphi$ and~$\vartheta$ to the right- and left-moving fields 
\beq
\varphi_\pm(x) = \frac{\,1}{\sqrt{K\,}}\,\varphi(x) \mp \sqrt{K\,}\vartheta(x),
\label{2.8}
\eeq
obeying the commutation relations \mbox{$\bigl[\partial_x\varphi_\pm,\varphi_\mp(y)\bigr] = 0$} 
and
\mbox{$\bigl[\partial_x\varphi_\pm,\varphi_\pm(y)\bigr] = \pm\msk 2\pi i{\mkern 0.5mu}\delta(x-y)$}. Both the total momentum \eq{2.2} and the Luttinger-liquid Hamiltonian~\eq{2.5} are chiral, i.e., diagonal in the basis of the right- and left-movers,
\beq
P = P_+ + P_-, 
\quad
H_0 = v(P_+ - P_-),
\label{2.9}
\eeq
where 
\beq
P_\pm  = \pm\,\frac{\hbar}{4\pi}\!\int\!dx\,(\partial_x\varphi_\pm)^2
\label{2.10}
\eeq 
are the momenta of the right- and left-moving excitations. 

The next-order contribution in the gradient expansion,~$H_1$ [see Eqs.~\eq{2.6} and \eq{2.7}], is the sum of the chiral part
\beq
\widetilde H_1 = \frac{\hbar^2}{12\pi m_\ast}\!\int_0^L\!dx
\bigl[(\partial_x\varphi_+)^3 + (\partial_x\varphi_-)^3 \bigr], 
\label{2.11}
\eeq
and a non-chiral part that includes integrals of $(\partial_x\varphi_\pm)^2(\partial_x\varphi_\mp)$. In \Eq{2.11}, $m_\ast$ has units of mass and satisfies \mbox{$m/m_\ast = (3/4)K^{-1/2}(1 +\beta K^2)$}. With $\beta$ given by \Eq{2.7}, this gives \Eq{1.3}. 

In order to see that $m_\ast$ indeed represents the effective mass of the elementary excitations, we focus on the chiral terms in $H_0 + H_1$, 
\begin{subequations}
\label{2.12}
\beqa
\widetilde H &=& H_+ + H_-, 
\label{2.12a}\\
H_\pm \!\! &=& \pm\, vP_\pm 
+ \frac{\hbar^2}{12\pi m_\ast}\!\int_0^L\! dx\msk(\partial_x\varphi_\pm)^3,~~~~~
\label{2.12b}
\eeqa
\end{subequations}
treating the remainder of the Hamiltonian as a perturbation. Because the four operators $P_\pm,H_\pm$ commute, it is sufficient to consider the right-movers only. Although the Hamiltonian $H_+$ is not quadratic, it can be diagonalized~\cite{Haldane_LL,Rozhkov} with the help of the well-known mapping~\cite{bosonization,Giamarchi,Haldane_LL} between one-dimensional bosons and fermions. In terms of such effective fermions, the right-moving bosons are described by the noninteracting Hamiltonian~\cite{Haldane_LL}
\begin{subequations}
\label{2.13}
\beqa
H_+ &=& vP_+ + \frac{\hbar^2}{2m_\ast}\!\int_0^L\!dx\,
\colon\!(\partial_x\psi)^\dagger (\partial _x\psi)\colon\nmsk,
\qquad
\label{2.13a} \\
P_+ &=& \!-\,i\hbar\!\int_0^L\!dx\,
\colon\nmsk\psi^\dagger(x)\msk\partial_x\psi(x)\colon\nmsk,
\label{2.13b}
\eeqa
\end{subequations}
and the boundary condition \mbox{$\varphi_+(x) = \varphi_+(x + L)$} translates to \mbox{$\delta N_+ = \int_0^L\!dx\msk\colon\!\psi^\dagger(x)\psi(x)\colon\! = 0$}. 
In Eqs.~\eq{2.13}, the fermionic field is given by \mbox{$\psi(x) = L^{-1/2} e^{\msk i\varphi_+(x)}$},   
with the right-hand side of this expression understood as being normal-ordered with respect to the bosonic vacuum. The colons in Eqs.~\eq{2.13} denote the normal ordering with respect to the fermionic vacuum in which all states with positive (negative) wave numbers are empty (occupied).

In the fermionic representation, construction of excitations is straightforward. We are interested in simultaneous eigenstates of $H_+$, $P_+$, and $\delta N_+$ with \mbox{$\delta N_+ = 0$}. Any such state is a superposition of the particle- and hole-type elementary excitations of the gas of effective fermions. The particle excitation is obtained by promoting a fermion from the Fermi level to one of the unoccupied single-particle states, whereas in the hole excitation a fermion is promoted from one of the occupied single-particle states to the Fermi level. Such particle (hole) excitations have the largest (smallest) possible energy for a given momentum, given by the first two terms in the expansions~\eq{1.1}.

Perturbation theory in $\delta H = H - \widetilde H$ transforms the right-moving eigenstate of $\widetilde H$ to the eigenstate of the full Hamiltonian $H$. 
This state is also an eigenstate of the total momentum~$P$ with the same eigenvalue. The leading contribution in $\delta H$ consists of dimension-3 operators $(\partial_x\varphi_\pm)^2(\partial_x\varphi_\mp)$. The key observation~\cite{Rozhkov} is that these operators are not chiral and have zero expectation value in the right-moving eigenstate of~$\widetilde H$. Accordingly, correction to the energy due to these operators arises only in the second order, yielding $\delta\varepsilon \propto p^3$, which does not affect the first two terms in the expansion \eq{1.1}.

\vspace{-1.5ex}
\section{Effective mass from Bethe ansatz}
\label{Bethe}

In this Section we derive low-energy excitation spectra for exactly solvable models and verify the validity of the phenomenological expressions \eq{1.1}-\eq{1.3}. Specifically, we consider a system of identical particles described by the Hamiltonian
\beq
H = \frac{\,\hbar^2}{2m\,}\!\left\lbrace
-\sum_{l}\frac{\,\partial^2}{\partial x_l^2} 
+ \sum_{\,l\,\neq\, l'}\!V(x_l - x_{l'}\!)
\right\rbrace
\label{3.1}
\eeq
in the thermodynamic limit when both the number of particles $N$ and the system size $L$ are taken to infinity with their ratio $n_0 = N/L$ kept fixed. Our consideration applies to models integrable by Bethe ansatz~\cite{Lieb,Yang,KBI,Sutherland_book}, including the Lieb-Liniger model~\cite{Lieb} describing bosons with contact repulsion
\beq
V(x) = c\msk\delta(x),
\label{3.2}
\eeq
and the hyperbolic Calogero-Sutherland model~\cite{Calogero,Sutherland} with the interaction potential given by  
\beq
V(x) = \frac{\lambda(\lambda - 1)}{a^2\sinh^2(x/a)}.
\label{3.3}
\eeq
(Because the potential \eq{3.3} is impenetrable, the statistics of the constituent particles is irrelevant.)  As mentioned in Sec.~\ref{intro}, excitation spectra in the limits \mbox{$0 < c/n_0\ll 1$} and $c/n_0 \gg 1$ for the Lieb-Liniger model \eq{3.2} and in the regime \mbox{$1\ll\exp(1/a n_0)\ll\lambda$} for the hyperbolic Calogero-Sutherland model \eq{3.3} have been studied analytically in~\cite{PM_LiebLiniger,PM_WignerToda,PM_KdV,Zoran} and found to be in agreement with Eqs.~\eq{1.1}-\eq{1.3}. The present consideration is applicable to all $c > 0$ in \Eq{3.2} and to all $\lambda > 0$ and any finite $a$ in \Eq{3.3}. 

In Bethe ansatz~[\onlinecite{Lieb,Yang,KBI,Sutherland_book}], many-body eigenstates are parametrized by sets of $N$ different rapidities $k_1, k_2,\ldots, k_{N}$. In this respect, the classification of eigenstates is analogous to that of noninteracting Fermi gas, in which any state is characterized by $N$ different wave numbers. The structure of the ground state is also similar to fermions: the ``occupied'' rapidities densely fill the interval $|k_i|\leq q$, where the ``Fermi rapidity'' $q$ is in one-to-one correspondence with the mean density $n_0$~\cite{Lieb,Yang,Sutherland_book,KBI}. The analogy extends to excited states, which can be viewed as superpositions of the particle- and hole-type excitations of the ``Fermi gas'' of rapidities, also known as, respectively, type~I and type II excitations~\cite{Lieb}. The momenta and energies of the right-moving $(p>0)$ excitations satisfy~\cite{Lieb,Yang,KBI,Sutherland_book}
\beq
p = 2\pi\hbar \left|\int_{q}^k\!dk'\rho(k',q)\right|,
\quad
\varepsilon = \left|\int_{q}^k\!dk'\sigma(k',q)\right|
\label{3.4}
\eeq
with $k > q~(|k| < q)$ for the particle (hole) excitation. The density of rapidities in the ground state \mbox{$\rho(k,q) = L^{-1}\!\sum_{i=1}^{N} \delta(k-k_i)$} depends on both $k$ and $q$ and obeys the Lieb equation~\cite{Lieb,KBI,Sutherland_book}
\beq
\rho(k,q) + \frac{1}{2\pi}\!\int_{-q}^{q}\!\!dk'{\mkern 1mu}\Theta'(k-k'){\mkern 1mu}\rho(k',q) 
= \frac{1}{2\pi}\msk,
\label{3.5} 
\eeq
in which $q$ enters as an independent parameter. The function $\sigma(k,q)$ in the second equation in~\eq{3.4} is the derivative of the energy function introduced in Ref.~\cite{Yang}. It obeys the Yang-Yang equation~\cite{Yang,KBI,Sutherland_book}
\beq
\sigma(k,q) + \frac{1}{2\pi}\!\int_{-q}^{q}\!\!dk' {\mkern 1mu}\Theta'(k - k'){\mkern 1mu}\sigma(k',q) 
= \frac{\hbar^2 k}{m}.
\label{3.6} 
\eeq
In Eqs.~\eq{3.5} and \eq{3.6}, $\Theta'(k)$ is the derivative of the two-particle scattering phase shift. The phase shift is given by \mbox{$\Theta(k) = - \,2\arctan(k/c)$} for the Lieb-Liniger model~\cite{Lieb,Sutherland_book,KBI} and by
\mbox{$\Theta(k) = 2{\mkern 1mu}\text{Im}\bigl[\ln\Gamma\!\left(\lambda + iak/2\right) - \ln\Gamma\!\left(1 + iak/2\right)\bigr]$} for the hyperbolic Calogero-Sutherland model~\cite{Sutherland,Sutherland_book}. 

Unlike the standard Calogero-Sutherland model corresponding to the infinite $a$ limit in \Eq{3.3}, the phase shifts in the cases we discuss are not singular. Accordingly, Eqs.~\eq{3.5} and \eq{3.6} yield analytic at all~$k$ functions \mbox{$\rho(k,q) = \rho(-k,q)$} and \mbox{$\sigma(k,q) = -\msk\sigma(-k,q)$}. Expanding $\rho$ and $\sigma$ in Eqs.~\eq{3.4} in Taylor series near $k = q$, we obtain \Eq{1.1} with
\beq
v = v(q,q) = \frac{\sigma_0}{2\pi\hbar\rho_0},
\quad
\frac{1\,}{\,m_\ast} = \frac{1\,}{2\pi\hbar\rho_0} \bigl(v^\prime_k\bigr )_{k \msk= \msk q},
\label{3.7}
\eeq
where 
\beq
v(k,q) = \frac{\sigma(k,q)}{2\pi\hbar\rho(k,q)}
\label{3.8}
\eeq
and $\rho_0 = \rho(q,q)$, $\sigma_0 = \sigma(q,q)$, $v^\prime_k = \partial v(k,q)/\partial k$.

\begin{widetext}
We now differentiate \Eq{3.5} with respect to $q$ and \Eq{3.6}  with respect to $k$. This gives
\begin{subequations}
\label{3.9}
\beqa
&&\rho^\prime_q(k,q) 
+ \frac{1}{2\pi}\!\int_{-q}^{q}\!\!dk'{\mkern 1mu}\Theta'(k-k'){\mkern 1mu}\rho^\prime_q(k',q) 
= -\msk\frac{\rho_0}{2\pi}
\bigl[\Theta'(k - q) + \Theta'(k+ q)\bigr],
\label{3.9a} \\
&& \sigma^\prime_k(k,q)
+ \frac{1}{2\pi}\!\int_{-q}^{q}\!\!dk' {\mkern 1mu}\Theta'(k - k'){\mkern 1mu}\sigma^\prime_{k'}(k',q) 
= \frac{\sigma_0}{2\pi}
\bigl[\Theta'(k - q) + \Theta'(k+ q)\bigr] + \frac{\msk\hbar^2}{m\msk}.
\label{3.9b}
\eeqa
\end{subequations}
\end{widetext}

\noindent
Comparison of \Eq{3.9b} with Eqs.~\eq{3.5} and \eq{3.9a} shows that $\sigma^\prime_k = (2\pi\hbar^2\nmsk/m)\rho - (\sigma_0/\rho_0)\rho^\prime_q$, or
\beq
\frac{\rho^\prime_q}{\rho_0} + \frac{\sigma^\prime_k}{\sigma_0}
=  \frac{2\pi\hbar^2}{m\sigma_0}\rho\msk.
\label{3.10}
\eeq
Similarly, differentiating Eqs.~\eq{3.5} and \eq{3.6} with respect to $k$ and~$q$ and comparing the resulting equations, we find
\beq
\frac{\rho^\prime_k}{\rho_0} + \frac{\sigma^\prime_q}{\sigma_0} = 0.
\label{3.11}
\eeq
Using Eqs.~\eq{3.8}, \eq{3.10} and \eq{3.11}, we obtain
\beq
\bigl(v^\prime_k - v^\prime_q\bigr )_{k \msk= \msk q} = \frac{\hbar}{m}\msk.
\label{3.12}
\eeq
This relation allows us to express the partial derivative $v^\prime_k$ in \Eq{3.7} via the total derivative
$dv/dq = \bigl(v^\prime_k + v^\prime_q\bigr )_{k \msk= \msk q}$, resulting in
\beq
\frac{1\,}{\,m_\ast} = \frac{1}{4\pi\hbar\rho_0}\nmsk\left(\frac{\hbar}{m} + \frac{dv}{dq}\right).
\label{3.13}
\eeq

In order to compare \Eq{3.13} with the phenomenological result \eq{1.3}, we need to eliminate the Fermi rapidity~$q$ in favor of the particle density $n_0$. The two are related via the normalization condition~\cite{Lieb,Sutherland_book,KBI}
\beq
n_0(q) = \int_{-q}^{q}\!\! dk\msk\rho(k,q),
\label{3.14} 
\eeq
which gives $dn_0/dq = 2\rho_0 + \int_{-q}^q\nmsk dk\msk \rho^\prime_q(k,q)$.
Substituting here~$\rho^\prime_q$ from \Eq{3.10}, we obtain
\beq
\frac{dn_0}{dq\,} = \frac{\msk\hbar n_0}{mv\msk}. 
\label{3.15}
\eeq
Combined with the formula~\cite{Sutherland_book,KBI} $d\mu/dq = \hbar v$, \Eq{3.15} gives $d\mu/dn_0 = (m/n_0){\mkern 1mu}v^2$, which turns the phenomenological expression for the sound velocity~\eq{1.2} into an identity~\cite{Lieb,Sutherland_book,KBI}.

Substituting $dv/dq = (dv/dn_0)(dn_0/dq)$ and taking into account \Eq{3.15}, we rewrite \Eq{3.13} as
\beq
\frac{\,m_\past}{\,m_\ast} = \frac{1}{4\pi{\mkern -1.25mu}\rho_0 v}\frac{d (vn_0)}{dn_0}\msk. 
\label{3.16}
\eeq
It remains to express $\rho_0$ here via the Luttinger-liquid parameter $K$. To this end, we multiply Eqs.~\eq{3.5} and \eq{3.9b} by $\sigma^\prime_k(k,q)$ and $\rho(k,q)$, respectively, and integrate over $k$ in the interval $-q<k<q$. The left-hand sides of the resulting equations are identical. Equating the right-hand sides and evaluating the integrals with the help of Eqs.~\eq{3.5} and~\eq{3.14}, we obtain
the formula~\cite{Sutherland_book,KBI,footnote} 
\beq
\rho_0\sigma_0 = \frac{\msk\hbar^2 n_0}{2m\msk}\msk.
\label{3.17}
\eeq
Comparison of the first equation in \eq{3.7} with \Eq{3.17} then yields the relation we seek, 
\beq
2\pi{\mkern -1.25mu}\rho_0 = \sqrt{K\msk}.
\label{3.18}
\eeq
Substituting \Eq{3.18} into \Eq{3.16}, we finally arrive at \Eq{1.3}.

\vspace{-2ex}
\section{Discussion}
\label{Discussion}

In this paper we considered two different routes to obtaining spectra of elementary excitations of Galilean-invariant one-dimensional quantum liquids. In Sec.~\ref{phenomenology} we discussed the particle (hole) excitations in the gas of effective noninteracting fermions that emerge as a result of the diagonalization of the phenomenological hydrodynamic Hamiltonian. In Sec.~\ref{Bethe} we studied Lieb's type~I (type~II) excitations in the Bethe ansatz formalism~\cite{models}. Both routes lead to Eqs.~\eq{1.1}-\eq{1.3} for the low-energy spectra. The agreement shows that Lieb's picture of elementary excitations as particles or holes of the ``Fermi gas'' of rapidities can be understood literally at low momenta, when the excitations are indeed particles or holes in the gas of effective fermions. 

Elementary excitations reveal themselves in the behavior of dynamic correlation functions, such as the dynamic structure factor (Fourier-transform of the density-density correlation function) $S(p,\varepsilon)$. In accordance with the phenomenological picture of weakly interacting effective fermions~\cite{Rozhkov}, at small momenta almost the entire spectral weight of $S(p,\varepsilon)$ is spread uniformly over the interval of energies \mbox{$\varepsilon_-(p) < \varepsilon <\varepsilon_+(p)$}. At $\varepsilon$ approaching $\varepsilon_\pm(p)$, the structure factor exhibits power-law singularities~\cite{PKKG,KPKG} characterized by exponents $\mu_\pm(p)$ that scale linearly at small $p$~\cite{PKKG}. 

It is natural to ask whether expansions similar to Eqs.~\eq{1.1}-\eq{1.3} for the excitation spectra can be derived for the exponents $\mu_\pm(p)$. It is easy to see that this is not the case. Indeed, the singularities arise due to interaction between the effective fermions~\cite{PKKG,KPKG,ISG,IG}. In addition, the exponents $\mu_\pm(p)$ are sensitive to the cubic and higher-order in~$p$ corrections to their spectra~\cite{ISG,IG}. At the level of the gradient expansion of the hydrodynamic Hamiltonian, the relevant contributions are represented by operators of scaling dimension 4 and higher. Some of these terms, such as the dimension-4 operators $(\partial^2_x\varphi_\pm)^2$, can not be written as products of powers of $\partial_x\varphi$ and $\partial_x\vartheta$. Accordingly, their coefficients can not be found by invoking the Galilean invariance and considering the response of the ground state to the variation of the mean density. Thus, it is not possible to obtain simple universal phenomenological expressions similar to Eqs.~\eq{1.2} and \eq{1.3} for the coefficients in the expansions of $\mu_\pm(p)$. By the same token, no such expressions exist for the coefficients of the cubic and higher-order terms in the expansions~\eq{1.1} of the excitation spectra.  

\vspace{-3ex}
\begin{acknowledgments} 
\vspace{-2ex}
We thank L. I. Glazman for discussions. This work was supported by the US Department of Energy, Office of Science, Materials Sciences and Engineering Division. We are grateful to the Aspen Center for Physics (NSF Grant No. PHYS-1066293) for hospitality.
\end{acknowledgments}

\vspace{-1ex}



\begin{thebibliography}{99}

\bibitem{PinesNozieres}
D. Pines and P. Nozi\`eres, 
\textit{The Theory of Quantum Liquids} (Westview Press, Boulder, 1999);
I. M. Khalatnikov, \textit{An Introduction to the Theory of Superfluidity}  
(Westview Press, Boulder, 2000).  

\bibitem{Giamarchi}
T. Giamarchi, \textit{Quantum Physics in One Dimension} (Clarendon Press, Oxford, 2004).

\bibitem{hydrodynamics}
V. N. Popov, Theor. Math. Phys. \textbf{11}, 478 (1972); 
V. N. Popov, Theor. Math. Phys. \textbf{11}, 565 (1972);
K. B. Efetov and A. I. Larkin, Sov. Phys. JETP \textbf{42}, 390 (1975);
F. D. M. Haldane, Phys. Rev. Lett. \textbf{47}, 1840 (1981).

\bibitem{Haldane_LL}
F. D. M. Haldane, J. Phys. C \textbf{14}, 2585 (1981).

\bibitem{Rozhkov}
A. V. Rozhkov, Eur. Phys. J. B \textbf{47}, 193 (2005).

\bibitem{Pereira}
R. G. Pereira, J. Sirker, J. -S. Caux, R. Hagemans, J. M. Maillet, S. R. White, and I. Affleck,
Phys. Rev. Lett. \textbf{96}, 257202 (2006);
J. Stat. Mech. (2007) P08022.

\bibitem{ISG}
A. Imambekov, T. L. Schmidt, and L. I. Glazman, 
Rev. Mod. Phys. \textbf{84}, 1253 (2012).

\bibitem{Sutherland_book}
B. Sutherland, \textit{Beautiful Models} (World Scientific, Singapore, 2004).

\bibitem{KBI} 
V. E. Korepin, N. M. Bogoliubov, and A. G. Izergin,
\textit{Quantum Inverse Scattering Method and Correlation Functions}
(Cambridge University Press, Cambridge, 1997).

\bibitem{Lieb} 
E. H. Lieb and W. Liniger, Phys. Rev. \textbf{130}, 1605 (1963);
E. H. Lieb, Phys. Rev. \textbf{130}, 1616 (1963).

\bibitem{Calogero}
F. Calogero, O. Ragnisco, and C. Marchioro, 
Lett. Nuovo Cimento \textbf{13}, 383 (1975);
F. Calogero, Lett. Nuovo Cimento \textbf{16}, 22 (1976).

\bibitem{Sutherland}
B. Sutherland, Rocky Mount. J. Math. \textbf{8}, 413 (1978).

\bibitem{PM_LiebLiniger} 
M. Pustilnik and K. A. Matveev, Phys. Rev. B \textbf{89}, 100504(R) (2014).

\bibitem{PM_KdV}
M. Pustilnik and K. A. Matveev, Phys. Rev. B \textbf{92}, 195146 (2015).

\bibitem{Zoran} 
Z. Ristivojevic, Phys. Rev. Lett. \textbf{113}, 015301 (2014).

\bibitem{AP}
G. E. Astrakharchik and L. P. Pitaevskii, Europhys. Lett. \textbf{102}, 30004 (2013).

\bibitem{PM_WignerToda}
M. Pustilnik and K. A. Matveev, Phys. Rev. B \textbf{91}, 165416 (2015).

\bibitem{qToda}
E. K. Sklyanin, Lecture Notes in Physics, \textbf{226}, 196 (1985); M. C. Gutzwiller, Ann. Phys., \textbf{133}, 304 (1981);
M. Toda, \textit{Theory of Nonlinear Lattices, 2nd Ed.} (Springer, Berlin, 1989).

\bibitem{MA} 
K. A. Matveev and A. V. Andreev, Phys. Rev. B \textbf{86}, 045136 (2012).

\bibitem{bosonization}
D. C. Mattis, J. Math. Phys. \textbf{15}, 609 (1974);
A. Luther and I. Peschel, Phys. Rev. B \textbf{9}, 2911 (1974);
S. Mandelstam, Phys. Rev. D \textbf{11}, 3026 (1975);
A. K. Pogrebkov and V. N. Sushko, Theor. Math. Phys. \textbf{24}, 935 (1975).

\bibitem{Yang}
C. N. Yang and C. P. Yang, J. Math. Phys. \textbf{10}, 1115 (1969).

\bibitem{footnote}
Compared to Refs.~\cite{Sutherland_book,KBI}, our derivation of \Eq{3.17} avoids explicit use of the formal inverse of the integral operator that appears in the left-hand sides of Eqs.~\eq{3.5}, \eq{3.6}, and \eq{3.9}.

\bibitem{models}
Our consideration is applicable to all models with excitation spectra given by Eqs.~\eq{3.4}-\eq{3.6} with nonsingular two-particle scattering phase shifts. 

\bibitem{PKKG}
M. Pustilnik, M. Khodas, A. Kamenev, and L. I. Glazman, 
Phys. Rev. Lett. \textbf{96}, 196405 (2006).
 
\bibitem{KPKG}
M. Khodas, M. Pustilnik, A. Kamenev, and L. I. Glazman, 
Phys. Rev. Lett. \textbf{99}, 110405 (2007).
 
\bibitem{IG} 
A. Imambekov and L. I. Glazman, 
Phys. Rev. Lett. \textbf{102}, 126405 (2009).
 
\end{thebibliography}
\end{document}